# Mechanism of preferential adsorption of SO$_2$ into two microporous paddle wheel frameworks M(bdc)(ted)$_{0.5}$

Kui Tan,[†] Pieremanuele Canepa,[‡] Qihan Gong,[§] Jian Liu,[∥] Daniel H. Johnson,[‡] Allison Dyevoich,[§] Praveen K. Thallapally,[∥] Timo Thonhauser,[‡] Jing Li,[§] and Yves J. Chabal*[†]

[†]Department of Materials Science & Engineering, University of Texas at Dallas, Richardson, Texas 75080
[‡]Department of Physics, Wake Forest University, Winston-Salem, North Carolina 27109
[§]Department of Chemistry and Chemical Biology, Rutgers University, Piscataway, New Jersey 08854
[∥]Energy and Environment Directorate, Pacific Northwest National Laboratory, Richland, Washington 99352

**ABSTRACT:** The selective adsorption of a corrosive gas, SO$_2$, into two microporous pillared paddle-wheel frameworks M(bdc)(ted)$_{0.5}$ [M = Ni, Zn; bdc = 1,4-benzenedicarboxylate; ted = triethylenediamine] is studied by volumetric adsorption measurements and a combination of *in-situ* infrared spectroscopy and *ab initio* density functional theory (DFT) calculations. The uptake of SO$_2$ in M(bdc)(ted)$_{0.5}$ at room temperature is quite significant, 9.97 mol/kg at 1.13 bar. The major adsorbed SO$_2$ molecules contributing to the isotherm measurements are characterized by stretching bands at 1326 and 1144 cm$^{-1}$. Theoretical calculations including van der Waals interactions (based on vdW-DF) suggest that two adsorption configurations are possible for these SO$_2$ molecules. One geometry involves an SO$_2$ molecule bonded through its sulfur atom to the oxygen atom of the paddle-wheel building unit and its two oxygen atoms to the C-H groups of the organic linkers by formation of hydrogen bonds. Such a configuration results in a distortion of the benzene rings, which is consistent with the experimentally observed shift of the ring deformation mode. In the other geometry, SO$_2$ establishes hydrogen bonding with -CH$_2$ group of the ted linker through its two oxygen atoms simultaneously. The vdW-DF-simulated frequency shifts of the SO$_2$ stretching bands in these two configurations are similar and in good agreement with spectroscopically measured values of physisorbed SO$_2$. In addition, the IR spectra reveal the presence of another minor species, characterized by stretching modes at 1242 and 1105 cm$^{-1}$ and causing significant perturbations of MOFs vibrational modes (CH$_x$ and carboxylate groups). This species is more strongly bound, requiring a higher temperature (~150 °C) to remove it than for the main physisorbed species. The adsorption configurations of SO$_2$ into M(bdc)(ted)$_{0.5}$ derived by infrared spectroscopy and vdW-DF calculations provide the initial understanding to develop microporous metal organic frameworks materials based on paddle-wheel secondary-building units for SO$_2$ removal in industrial processes.

## 1. Introduction

Metal organic frameworks (MOFs) are a new class of solid porous materials that are attracting great interest as a potential material for gas storage and separation because of their extraordinary surface areas, and fine tunable surface properties.[1-3] Until now, MOFs have been widely investigated for the storage and separation of H$_2$, CH$_4$, CO$_2$ and hydrocarbons.[3-9] Studies to explore the use for MOFs for harmful gases (e.g. SO$_2$, H$_2$S, NO) adsorption and removal are relatively sparse compared to studies on their use for hydrogen storage and carbon capture.[10-13] In the past decades, adsorption of SO$_2$ into porous materials such as porous carbon, zeolites has been proposed and investigated by different techniques in order to develop industrial desulfurization process.[14-19] Recently, Zn-MOF-74 has been found to improve the SO$_2$ adsorption by a factor of 3 compared to activated carbons and be the best performing MOF among six prototypical MOFs.[20] Very recent breakthrough measurements showed that Mg-MOF-74 provides a better adsorption of SO$_2$ gas than other MOF-74 series (with Zn, Ni, Co).[21] Indeed, MOF-74 [M$_2$(dhtp), dhtp = 2,5-dihydroxyterephthalate] compounds have hexagonal one dimensional pores containing a high density of coordinatively unsaturated open metal sites that can potentially interact with guest molecules (see Figure S1 in Supporting information). Usually the generation of MOF structures with exposed metal sites such as MOF-74 compounds is regarded as an effective means to enhance their affinity toward gases.[3,4] More recently, there has been an effort to develop structures that have cooperative multiple interactions with guest molecules that are believed to improve the adsorption capacity.[22-24] For example, Schröder and coworkers reported a new non-amine-containing MOF complex called NOTT-300, which selectively adsorbs SO$_2$ to a level of 8.1 mmol/g at ambient temperature and 1.0 bar. The highlight of this finding is that OH group and C-H groups of benzene cooperatively interact with guest SO$_2$ molecules.[24]

To develop the understanding of MOF-based materials for SO$_2$ removal, we examined SO$_2$ adsorption in a prototypical metal organic framework M(bdc)(ted)$_{0.5}$ [M= Ni, Zn; H$_2$bdc = 1,4-benzenedicarboxylic; ted = triethylenediamine] and characterized the guest-host interaction by *in-situ* infrared spectroscopy and isotherm measurements. Our results show that high SO$_2$ capacity uptake can be achieved even when using coordinatively saturated MOFs M(bdc)(ted)$_{0.5}$, at levels up to

9.97 mmol/g at room temperature and 1.13 bar. Combining infrared spectroscopy and first principles calculations, the adsorption mechanism of $SO_2$ within this class of MOFs is explored. Multipoint interactions between $SO_2$ molecules and MOFs are found as possible binding sites within the frameworks: $SO_2$ interacting i) with the O site of the paddlewheel building units through its sulfur atom, and ii) with the C-H$_x$ groups on the organic linkers through its oxygen atoms.

## 2. Experimental section:

*Synthesis*: A mixture of nickel (II) chloride hexahydrate (0.107 g, 0.45 mmol), H$_2$bdc (0.060 g, 0.36 mmol), ted (0.033 g, 0.29 mmol) and 15 ml of DMF were transferred to Teflon-lined autoclave and heated at 120 °C for 2 days. Green crystalline powder of Ni(bdc)(ted)$_{0.5}$ was isolated by filtering and washed three times with 10 ml of DMF. (Yield: 83.8% based on the metal) Then the sample was heated at 120 °C again under a flow of dry N$_2$ for one day to remove the guest DMF molecules.

A mixture of zinc (II) nitrate hexahydrate (0.245 g, 0.82 mmol), H$_2$bdc (0.136 g, 0.97 mmol), ted (0.073 g, 0.65 mmol) and 15 ml of DMF were transferred to Teflon-lined autoclave and heated at 120 °C for 2 days. Cubic colorless crystals of Zn(bdc)(ted)$_{0.5}$ were isolated by filtering and washed three times with 10 ml of DMF. (Yield: 84.5% based on the metal) Then the sample was heated at 120 °C again under a flow of dry N$_2$ for one day to remove the guest DMF molecules.

*Infrared spectroscopy*: A powder of M(bdc)(ted)$_{0.5}$ (~2 mg) was pressed onto a KBr pellet(~1 cm diameter, 1-2 mm thick) and placed into a high pressure high temperature cell purchased from Specac (product number P/N 5850c) at the focal point of the sample compartment of the infrared spectrometer (Nicolet 6700, Thermo Scientific) equipped with a liquid N$_2$–cooled MCT-B detector. The cell was connected to a vacuum line for evacuation. $SO_2$ gas was introduced into the pressure cell and all spectra were recorded in transmission between 400 and 4000 cm$^{-1}$ (4 cm$^{-1}$ spectral resolution). The pressure was determined by calibrating the $SO_2$ gas IR absorption at low pressures (3 Torr) with a convection gauge (KJL 285801, KurtJLeskerCo) and then using the integrated IR absorption to determine the higher pressures since there is no dependence of the internal mode dipole moment on pressure.

*Adsorption isotherm*: Measurements of $SO_2$ uptake were performed using a volumetric system specially constructed for low-pressure experiments (see Figure S12). It comprises two chambers, A and B, whose volumes have been determined using helium. Chamber B is loaded with the sample and chamber A is a gas reservoir. A known amount of powder sample was placed in chamber B (volume = 4.134 cm$^3$) with a layer of glass wool on top to prevent sudden suction of the powder sample. The sample was activated at 150 °C for 12h with dynamic vacuum before each measurement. The activated sample weight was measured after experiment and was used to calculate the $SO_2$ adsorption results. The activated sample weight for Ni(bdc)(ted)$_{0.5}$, Zn(bdc)(ted)$_{0.5}$ and Mg-MOF-74 are 75.5 mg, 94.2 mg and 95.6 mg respectively. The activated sample was sealed in chamber B under vacuum. $SO_2$ (the total impurity of $SO_2$ gas is less than 0.1% and the water concentration is less than 0.01%) was introduced into chamber A and then the valve connecting chamber A and B was opened to let the gas adsorption reach equilibrium. This step was repeated as necessary to obtain the adsorption data points. As for the desorption part, a similar but reverse procedure was adopted. The chamber A was evacuated (1×10$^{-4}$ Torr) and then the valve connecting chamber A and B was opened to let the gas desorption reach equilibrium. This step was also repeated as necessary to obtain all desorption data points. The criterion for equilibrium was that the change in pressure was less than 0.1% within 15 min. A longer equilibrium time (1h) was used but no significant difference was observed for the pressure readings. The $SO_2$ quantity change was determined by the pressure change in chamber A and B. The recorded pressures were used to calculate the adsorbed amount based on basic mass balance equations. An ideal gas law was adopted and the volume of the sample was estimated to be 0.15 cm$^3$, and need to be subtracted from the volume of chamber A in the calculations. The non-ideality of $SO_2$ was accounted by using the compressibility factor of $SO_2$, which is 0.99 under our experiment conditions. The activation procedure described above was repeated before proceeding to the next cycle of $SO_2$ adsorption and desorption measurements. Based on our experimental observations, the Ni(bdc)(ted)$_{0.5}$ can maintain its $SO_2$ capacity after at least five full cycles of adsorption and desorption.

*DFT calculations*: The theoretical results presented in this manuscript were obtained using the van der Waals exchange and correlation functional vdW-DF[25,26] as implemented in the Vienna *ab initio* Simulation Package, VASP[27-30]. This functional has already been successfully applied to a number of similar studies, investigating small molecule adsorption in MOFs[31-35]. vdW-DF is able to accurately describe the weak van der Waals forces originating from non-local interactions. We simulated the loading and the co-adsorption of $SO_2$ in M(bdc)(ted)$_{0.5}$, with M= Ni, Zn. Projector augmented-wave (PAW)[36,37] theory combined with a plane-wave cutoff of 700 eV was used to describe the wave functions. The convergence threshold for the total energy was set to 1*10$^{-8}$ eV, ensuring an accurate sampling of the complex potential energy surface of these MOFs. The structure of M(bdc)(ted)$_{0.5}$ interacting with $SO_2$, was relaxed using vdW-DF until the force criterion of 1*10$^{-4}$ eV·Å$^{-1}$, was satisfied. Experimentally the Zn(bdc)(ted)$_{0.5}$ structure exhibits some proton statistical disorder (previously studied by Kong *et al.*[38]) resulting in a tetragonal cell with $a$, $b$ = 10.93 Å and $c$ = 9.61 Å.

The Ni(bdc)(ted)$_{0.5}$ structure was obtained after full relaxation of the Zn(bdc)(ted)$_{0.5}$ where Zn atoms were replaced by Ni atoms. The optimized lattice constants for Ni(bdc)(ted)$_{0.5}$ are $a$ = b = 11.15 Å and $c$ = 9.53 Å, respectively. For Ni(bdc)(ted)$_{0.5}$ the magnetic moment exhibited by the Ni atoms (2 per cell) is approximately 1 μB Ni$^{-1}$ and remains entirely localized on the Ni species.

With these settings we fully relaxed the coordinates of the MOF and the adsorbed species, for which the adsorption energies are calculated. Starting from the well-converged adsorption geometries, we simulated the IR/Raman vibrational modes of $SO_2$. Vibrational frequencies were obtained by diagonalizing the dynamical matrix at the Γ point using a finite difference approach with a three-point formula and a calibrated displacement of 0.005 Å. Bader charges were computed using the fast implementation of Henkelman and coworkers.[39] The adsorption models and the graphic manipulation were carried out using J-ICE. [40]

## 3. Experimental results.

### 3.1 Adsorption isotherms.

The M(bdc)(ted)$_{0.5}$ series contains secondary-building units SBUs of two 5-coordinate metal cations bridged in a paddle wheel-type configuration (see Figure 1).[41-44] Compared to the MOFs with coordinatively unsaturated metal sites, M(bdc)(ted)$_{0.5}$ is more readily activated and highly porous (61.3%) with large BET surface areas exceeding 1700m$^2$/g.[44,45]

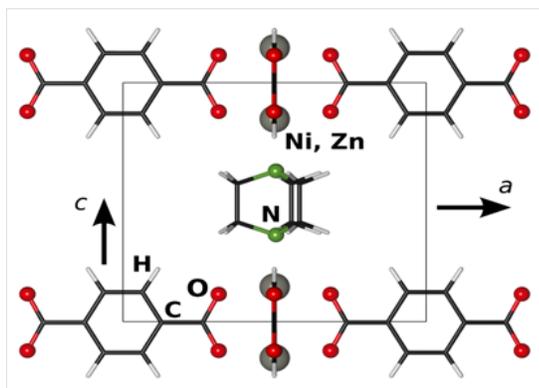

**Figure 1**. Structure of M(bdc)(ted)$_{0.5}$ (M=Ni and Zn) viewed along the *a* and *c* axes. Each paddle wheel building unit is linked by bdc within the layer of 2D network (xy plane). The apical sites of metal ions in the building units are bonded by ted molecules to generate the 3D porous frameworks. The coordinative bonds between metal ions Ni$^{2+}$, Zn$^{2+}$ and O, N atoms of bdc, ted linkers are omitted.

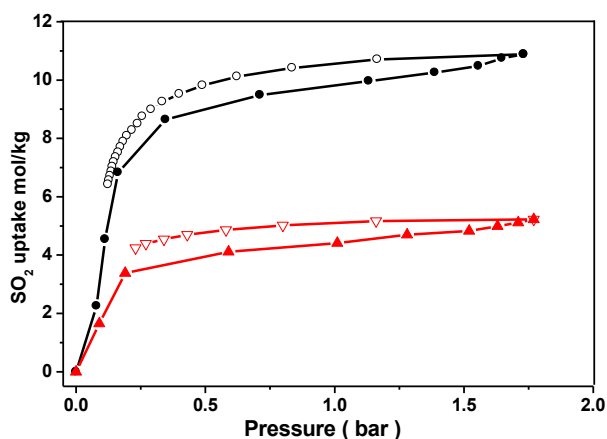

**Figure 2**. Adsorption isotherm of $SO_2$ in Ni(bdc)(ted)$_{0.5}$ (black), Zn(bdc)(ted)$_{0.5}$ (red) at room temperature for pressures up to ~1.8 bar. Solid symbol: adsorption; Empty symbol: desorption.

The $SO_2$ adsorption isotherms for activated Ni(bdc)(ted)$_{0.5}$ and Zn(bdc)(ted)$_{0.5}$ were collected at 298 K. The data are shown in Figure 2. For the pressures below 0.35 bar, the $SO_2$ uptake in Ni(bdc)(ted)$_{0.5}$ increases rapidly to 8.64 mol/kg, and then reaches 10.88 mol/kg at 1.73 bar. This is quite similar to the $SO_2$ isotherms observed in NOTT-300.[24] The rapid increase observed can be attributed to the strong interaction of $SO_2$ with the surface of the frameworks determined by spectroscopic measurements and theoretical calculations in a later section. Table 1 shows that at room temperature and around 1 bar, Ni(bdc)(ted)$_{0.5}$ outperforms the other reported metal organic frameworks materials in $SO_2$ adsorption with 9.97 mol/kg $SO_2$ uptake capacity, compared for instance with Mg-MOF-74, as previously shown in breakthrough measurements, found here under isotherm conditions to have 8.60 mol/kg $SO_2$ uptake capacity. In contrast, at low pressure (< 0.20 bar), MOF-74 and NOTT-300 adsorbs more $SO_2$ than Ni(bdc)(ted)$_{0.5}$. Comparing the uptake of different gases ($CO_2$, $N_2$, $CH_4$, $H_2$),[44,46,47] as shown in Table S2 and co-adsorption measurement in Supporting Information, it is found that Ni(bdc)(ted)$_{0.5}$ exhibits a preferential adsorption towards $SO_2$. The observed hysteresis during the desorption process is attributed to kinetic limitations. The $SO_2$ adsorption uptake in the isostructural Zn(bdc)(ted)$_{0.5}$ compound is much lower than in Ni(bdc)(ted)$_{0.5}$, which can be explained by the stability of M(bdc)(ted)$_{0.5}$ [M=Ni, Zn] under $SO_2$. The XRD pattern (see Figure S3) reveals that that the crystal structure of Ni(bdc)(ted)$_{0.5}$ is stable under $SO_2$ loading while that of Zn(bdc)(ted)$_{0.5}$ shows evidence of partial decomposition. The high thermal stability of Ni(bdc)(ted)$_{0.5}$ (over 400 ˚C as shown in Figure S4) makes it suitable for real world applications.

**Table 1.** Comparison of $SO_2$ adsorption capacities (uptakes in mol kg$^{-1}$) in selected MOF at room temperature. Pressure (Press) in bar.

| MOF | Uptake | Press | Temp | Reference |
| --- | --- | --- | --- | --- |
| Ni(bdc)(ted)$_{0.5}$ | 9.97 | 1.13 | 298 | This work |
|  | 4.54 | 0.11 | 298 | This work |
| Zn(bdc)(ted)$_{0.5}$ | 4.41 | 1.01 | 298 | This work |
| Mg-MOF-74 | 8.60 | 1.02 | 298 | This work |
|  | 6.44 | 0.11 | 298 | This work |
|  | 1.94 | a | 293 | 21 |
| NOTT-300 | 8.1 | 1 | 273 | 24 |
| $M_3[Co(CN)_6]_2$ | 2.5 | 1 | 298 | 13 |
| FMOF-2 | 2.19 | 1 | 298 | 12 |

[a] The dynamic capacity of $SO_2$ in Mg-MOF-74 measured by breakthrough measurement under 1000 mg m$^{-3}$ of the feed concentration.

### 3.2 Spectroscopic characterization of $SO_2$ adsorption and desorption in M(bdc)(ted)$_{0.5}$.

To unravel the nature of the interactions between $SO_2$ molecules and the Ni(bdc)(ted)$_{0.5}$ framework upon $SO_2$ adsorption, we performed *in-situ* IR measurements of $SO_2$ adsorption at room temperature as a function of pressure. Since the $SO_2$ gas IR absorption at pressures above 100 Torr is prohibitively high, a different procedure was employed to study the loadings above 45 Torr (see Figure S6). The activated sample was exposed to the SO$_2$ gas and the pressure cell quickly evacuated. The IR spectra were then recorded immediately after evacuation (acquisition time = 16 seconds). The adsorption state could be probed because the kinetics of SO$_2$ removal are slow enough, as shown in Figure 3. Several SO$_2$ associated bands, increasing as a function of initial pressure, can be identified in the IR spectra for Zn(bdc)(ted)$_{0.5}$ of Figure 3 and Figure S7. Two sharp bands observed at 1326 cm$^{-1}$ and 1144 cm$^{-1}$, exhibit a -36 cm$^{-1}$ and -7 cm$^{-1}$ red-shift, respectively (from the unperturbed values of 1362 cm$^{-1}$ and 1151 cm$^{-1}$ for the asymmetric and symmetric S=O stretch modes of SO$_2$ molecules).[48-50] Although these frequency shifts are typical of *physisorbed* SO$_2$ species,[49,51] we will show in section 4.1 that there can be charge transfer with the MOF.

A combination band, $\nu_{as}+\nu_s$, is observed at 2462 cm$^{-1}$. Two other bands appear at 1242 cm$^{-1}$ and 1105 cm$^{-1}$, most easily seen in the low pressure range (see inset of Figure 3), which are substantially more red-shifted compared to the gas phase values (-120 and -53 cm$^{-1}$, respectively) than these of the previous bands discussed above. SO$_2$ adsorption also induces significant changes to the frameworks vibrational modes: (1) a blue shift of the $\nu_{as}$(COO) mode from 1639 cm$^{-1}$ to 1695 cm$^{-1}$ ($\Delta\nu$ = +56 cm$^{-1}$) and of the $\nu_s$(COO) mode from 1433 cm$^{-1}$ to 1472 cm$^{-1}$ ($\Delta\nu$ = +39 cm$^{-1}$), and (2) a decrease in intensity of $\nu_{as}$(CH$_2$), $\nu_s$(CH$_2$) and $\nu$(CH) modes at 2874 cm$^{-1}$, 2938 cm$^{-1}$, and 3076 cm$^{-1}$, respectively. Furthermore, the CH$_2$ rocking mode and benzene ring deformation mode $\sigma_{12}$ are red-shifted by ~-8 cm$^{-1}$ (from 830 and to 822 cm$^{-1}$) and ~-12 cm$^{-1}$ (from 744 cm$^{-1}$ to 732 cm$^{-1}$). The $\delta$(COO) mode, initially at 810 cm$^{-1}$ shifts to 778 cm$^{-1}$ ($\Delta\nu$ = -32 cm$^{-1}$). These changes increase with SO$_2$ adsorption. The frameworks vibrational modes assignment is summarized in Table S3.

Figure 4 shows the evolution of SO$_2$ in Ni(bdc)(ted)$_{0.5}$ after a loading at 286 Torr. Upon evacuation at room temperature, the physisorbed species characterized by modes at 1326 cm$^{-1}$ and 1144 cm$^{-1}$ gradually desorbs. In contrast, the two bands at 1242 cm$^{-1}$ and 1105 cm$^{-1}$ remain relatively strong, indicating that another more strongly bound SO$_2$ species is present in the frameworks.

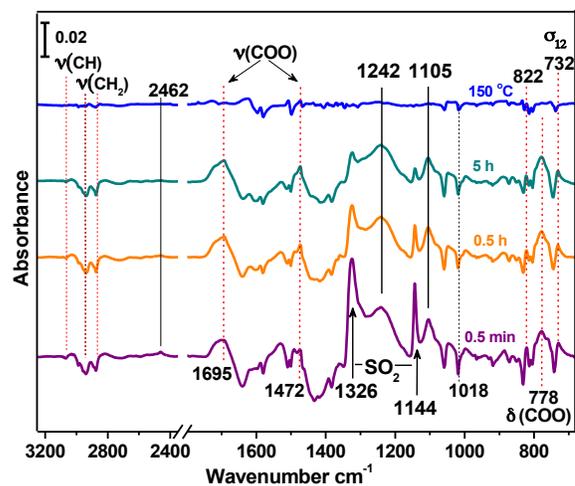

**Figure 4.** IR absorption spectra recorded 0.5 min, 0.5h and 5h after evacuation at room temperature after loading SO$_2$ at 286 Torr, and then after at 150 °C for additional 3 h. All spectra are referenced to the activated (i.e. empty) MOF. The vertical black lines are associated with SO$_2$-related features and the red dash or dot lines with MOF-related features arising from SO$_2$-induced perturbations.

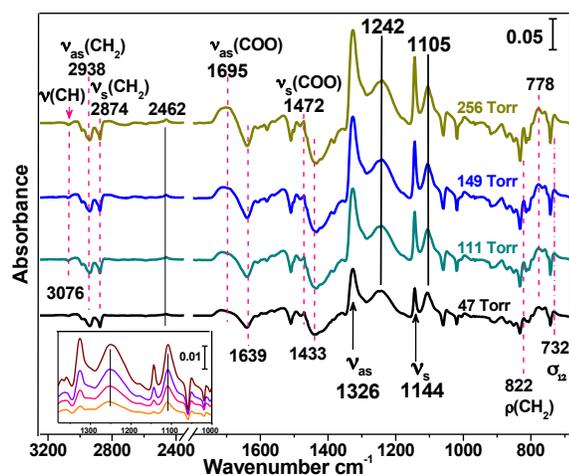

**Figure 3.** IR absorption spectra of SO$_2$ adsorption into Ni(bdc)(ted)$_{0.5}$ as a function of SO$_2$ initial pressure recorded immediately after evacuation (within 16 s). All spectra are referenced to the activated (i.e. empty) MOF. Inset shows the low pressure region from bottom to top: 600 mtor, 1 Torr, 2 Torr, 3 Torr. The black lines are associated with SO$_2$-related features and the red dash lines with MOF-related features arising from SO$_2$ induced perturbation.

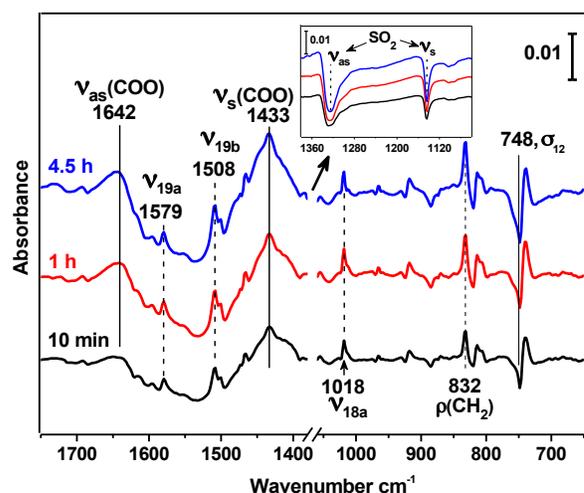

**Figure 5.** Differential spectra recorded during evacuation. The reference for all spectra is obtained at the very beginning of the evacuation process (i.e. within 0.5 min). The inset shows the physisorbed SO$_2$ in the frequency region of 1050 to 1380 cm$^{-1}$.

In co-adsorption experiments of CO$_2$ and SO$_2$ in Figure S9, we found that the presence of this other SO$_2$ species can decrease the CO$_2$ adsorption. The complete removal of this species required heating to 150 °C, as shown in the top spectrum in Figure 4. In the differential spectra, the perturbed benzene ring stretching bands $\nu_{18a}$, $\nu_{19b}$, $\nu_{19a}$ modes, benzene ring deformation modes $\sigma_{12}$, $\nu(COO)$, $\rho(CH_2)$ of ted linkers are partially recovered after the removal of the dominant, less strongly bound $SO_2$ molecules upon evacuation at room temperature. Comparison of Figures 4 and 5 indicates that the weakly bound $SO_2$ (characterized by modes at 1326 cm$^{-1}$ and 1144 cm$^{-1}$) causes the $\sigma_{12}$ mode to blue shift by 4 cm$^{-1}$ while the more strongly bound $SO_2$ characterized by modes at 1242 cm$^{-1}$ and 1105 cm$^{-1}$ causes the $\sigma_{12}$ mode red shift to 732 cm$^{-1}$. The perturbation of the $\nu(COO)$, $\nu(CH_2)$, $\delta(COO)$ bands and the benzene deformation mode $\sigma_{12}$ is removed after this latter species, characterized by modes at 1242 cm$^{-1}$ and 1105 cm$^{-1}$, has desorbed at elevated temperatures.

### 3.3 Perturbation with $D_2O$, $SO_2$ and $CO_2$:

To elucidate the nature of the perturbation of the MOF structure (i.e. phonons), we examined the effect of adsorption of *other* molecules (water and carbon dioxide) into the Ni(bdc)(ted)$_{0.5}$ frameworks on the IR absorption spectra (Figure 6). For water, $D_2O$ is used instead of $H_2O$ to avoid the interference of the $\beta(D_2O)$ (scissor) mode in the range of 1600 to 1700 cm$^{-1}$. We find that the perturbation effects induced by $SO_2$ adsorption are dependent on the preadsorbed gas and significantly different after incorporation of water molecules into the M(bdc)(ted)$_{0.5}$ frameworks. As previously observed and discussed,[35] the $\nu_{as}(COO)$ and $\nu_s(COO)$ modes, initially at ~1620 cm$^{-1}$ and ~1430 cm$^{-1}$, red shift to 1567 cm$^{-1}$ and 1365 cm$^{-1}$ upon introduction of the water molecules. The red shift is due to hydrogen bonding between $D_2O$ and the carboxylate group in the paddlewheel building units. In the case of $CO_2$, there is little perturbation of the host, which is consistent with the fact that $CO_2$, is known to be weakly physically absorbed into M(bdc)(ted)$_{0.5}$.[51] The response of the benzene ring deformation mode $\sigma_{12}$, $\delta(COO)$, is specific to each guest-molecule adsorption into the frameworks: $\sigma_{12}$ blue shift to 756 cm$^{-1}$ upon exposure to water vapors; $SO_2$ adsorption causes a more pronounced red shift of the $\delta(COO)$ mode than water molecules.

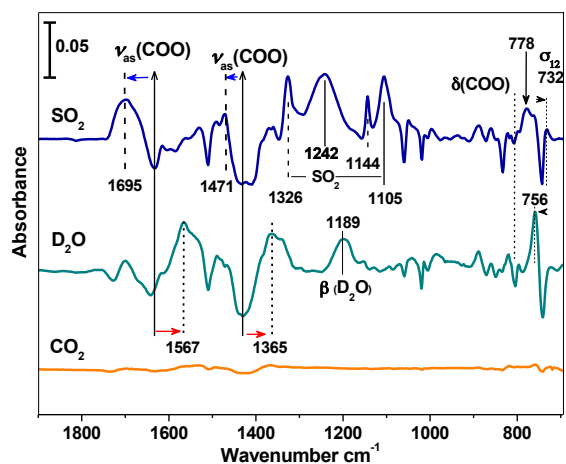

**Figure 6.** Perturbation of MOFs vibrational modes: IR absorption spectra of $SO_2$, $D_2O$, $CO_2$ adsorption into Ni(bdc)(ted)$_{0.5}$ recorded right after exposure following (a) 16 Torr $SO_2$, (b) 6 Torr $D_2O$, and (c) 16 Torr $CO_2$. All spectra are referenced to the activated (i.e. empty) MOF.

## 4. Discussion:

The infrared results indicate that $SO_2$ adsorbs into the M(bdc)(ted)$_{0.5}$ frameworks in two different ways, distinguished by their frequency shifts, effect on the host spectrum and bonding strength. From the desorption isotherms shown in Figure 2 and the IR absorption spectra in Figure 4, we conclude that the major species is weakly adsorbed $SO_2$, characterized by modes at 1326 cm$^{-1}$ and 1144 cm$^{-1}$. In addition, we observe that some minor species persists in the framework upon evacuation and can block access to other molecules such as $CO_2$ as indicated in Figure 4 and Figure S9. Upon the removal of this more strongly bound $SO_2$ species, the $CO_2$ uptake can be recovered, confirming that these $SO_2$ species inhibit the adsorption of incoming $CO_2$ molecules.

To determine the origin of all adsorbed species, we first turn to the literature. Previous experimental and theoretical studies of $SO_2$ interaction with metal oxide such as MgO and ZnO suggest that $SO_2$ readily reacts with exposed O sites at surfaces to form $SO_3$ and in some cases $SO_4$ species.[52-54] According to this picture, it is likely that $SO_2$ molecules preferentially interact with the oxygen site of the paddlewheel building units by Lewis acid-base interaction. For such an interaction, perturbations of the $\nu_s(COO)$ bands are expected, as observed in the IR spectra (see Sec. 3.2). However, other sites may also be occupied leading to other types of interactions. For instance, the "strong" perturbations of the linker bands such as $\nu(CH_x)$, $\nu_{18a}$, $\varrho(CH_2)$ and $\sigma_{12}$ indicate the possible interaction of $SO_2$ molecules with the benzene rings and the ted moieties in the frameworks as well.

While the modes associated with the dominant species at 1326 cm$^{-1}$ and 1144 cm$^{-1}$ correspond to physisorbed $SO_2$ according to the isotherms, the other two bands at 1242 cm$^{-1}$ and 1105 cm$^{-1}$ are clearly associated with more strongly bound $SO_2$ species, with frequencies more similar to previously observed values for sulfate species and chemisorbed $SO_3$ on MgO surfaces.[50,55,56] It is therefore tempting to suggest that strong Lewis adducts $SO_2$-carboxilate are formed within the Ni(bdc)(ted)$_{0.5}$ framework by a so-called $\eta^1$-S bridge configuration,[52,54,56] wherein $SO_2$ is attached via the S atom to two oxygen atoms of the paddle wheel building units, as shown in Figure S11. However, spectroscopic results in Figure 3 and 4 also suggest the possibility of $SO_2$ interaction with the ted organic linker since the appearance of the bands at 1242 and 1105 cm$^{-1}$ is correlated with a decrease in intensity of $\nu_{as}(CH_2)$, $\nu_s(CH_2)$ modes of the ted group.

To determine whether this interaction involves the ted group, two other isotypical metal organic frameworks without ted linker, but based on the same paddlewheel building units Zn(ndc)(bpee)$_{0.5}$, and Zn(ndc)(bpy)$_{0.5}$ were studied, where [ndc=2,6-naphthalenedicarboxylate, bpee=1,2-bis(4-pyridyl)ethylene and bpy=bipyridine](see Figure 7).[57,58] For these MOFs, bpee and bpy link the two-dimensional networks Zn$_2$(ndc)$_2$ to generate 3D pillared structures similar to Zn(bdc)(ted)$_{0.5}$. Since the bpy and bpee linkers are planar linkers in contrast to the 3-D ted linker, they are less bulky. They therefore would make it easier for $SO_2$ molecules to interact with the paddlewheel units, if the formation of a stable

Lewis SO$_2$-carboxilate species were in fact stable, in Zn(ndc)(bpee)$_{0.5}$, and Zn(ndc)(bpy)$_{0.5}$. However, no feature associated with strongly bound species could not be observed in the spectra of SO$_2$-loaded Zn(ndc)(bpee)$_{0.5}$, and Zn(ndc)(bpy)$_{0.5}$. Therefore, the formation of this stable Lewis-adduct (with O-O distance ~2.4-2.5 Å) do not appear to be favorable in paddlewheel units, possibly due to large O-O distances (~2.7- 2.8Å) in the carboxylate groups of the unit.[41,42,59]

These results suggest that interaction with the ted linker may be strong enough to induce the shifts observed in the SO$_2$ vibrational spectrum, 1242 and 1105 cm$^{-1}$, and to lead to a higher binding energy, as shown in Figure 4 and Figure 7. We have therefore examined such possibilities using first principles calculations that are useful to determine the binding energies associated with selected binding sites of SO$_2$ molecules within the Ni/Zn(bdc)(ted)$_{0.5}$, and to calculate the associated vibrational spectrum of both SO$_2$ and the MOF ligand modes, as described in section 4.1.

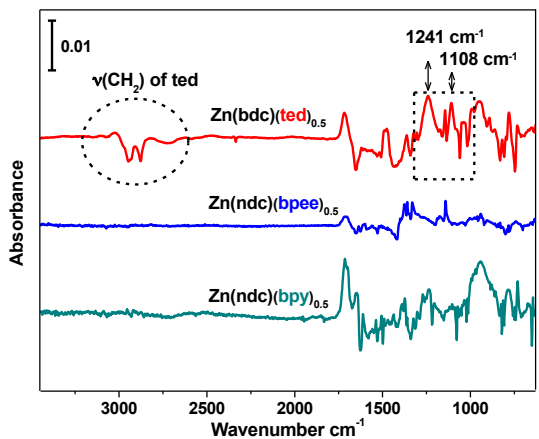

**Figure 7.** IR absorption spectra of SO$_2$ adsorption into activated Zn(bdc)(ted)$_{0.5}$, Zn(ndc)(bpee)$_{0.5}$ and Zn(ndc)(bpy)$_{0.5}$ at the pressure of 23 Torr, 20 Torr, 24 Torr recorded immediately after evacuation (within 16 s). All spectra are referenced to the activated (i.e. empty) MOF.

Ni/Zn(bdc)(ted)$_{0.5}$ offer several potential binding sites associated with the nature of their building block. However their potential energy surface (PES) is complex, presenting many adsorptions "pockets". To guide the construction of initial and realistic adsorption geometries, we initially assumed that the interaction of SO$_2$ with MOFs structure was similar to what occurs when SO$_2$ is adsorbed on simpler systems such MgO and ZnO surfaces.[52,54] This assumption is justified by the fact that oxygen atoms surrounding the metals sites of MOFs M(bdc)(ted)$_{0.5}$ "recreate" the chemical environment found on these surfaces. Theoretical and experimental evidence of the adsorption of SO$_2$ on MgO and ZnO surfaces suggests that SO$_2$ interacts strongly with its sulfur atom with the oxygen sites exposed at the surface, leading us to design an initial model of interaction where the oxygen atoms of SO$_2$ bind with the oxygen atoms of the bdc units (model SO$_2$-O-bdc). However, due to the complexity of PES of the M(bdc)(ted)$_{0.5}$ we also considered three other possible initial adsorption sites, based on the complementary electrostatic principle of Lewis-donors and acceptors. In general, S in SO$_2$ acts as an acceptor, and the functionalities on the linker as the donors. This has led us to select adsorption geometries (i.e. models) where SO$_2$ is in contact with *i)* the benzene ring of the bcd units (model SO$_2$-benzene), *ii)* with the nitrogen atom of the ted group (model SO$_2$-N-ted), and *iii)* the C-H groups of the ted linkers by establishing strong hydrogen bond (SO---H-C), which can be done with several such C-H groups that fully accessible in the MOF channel of the model SO$_2$-CH$_2$-ted system.

**4.1 Adsorption energy and charge analysis**

The fundamental quantity that governs the adsorption properties of SO$_2$ within the MOF is the adsorption energy ΔE, referred to thereafter as binding energy:

$$\Delta E = E_{MOF+G} - E_{MOF} - E_G \qquad (1)$$

Here E$_{MOF+G}$, E$_G$ and E$_{MOF}$ are the total energies of the fully relaxed MOF + SO$_2$, MOF alone, and SO$_2$ in gas phase, respectively. Knowledge of the vibrational frequencies (Sec. 4.2) of these systems becomes useful in the calculation of the zero-point energy (ZPE), ΔE$_{ZPE}$, the thermal correction ΔH(T), and entropy contribution ΔS. ΔH(T) allows for a more accurate comparison to measured heats of adsorption and is calculated considering the vibrational contribution (taking into account the rotational and translational degrees of freedom for the molecule in gas phase) to the adsorption energy.

We begin by discussing the binding energies and other relevant quantities for mono-adsorption cases of SO$_2$ molecules within Zn(bdc)(ted)$_{0.5}$ framework as detailed above and Figure 8 shows these initial alternative adsorption models.

The adsorption of SO$_2$ molecules on the groups presented by the MOF follow the trend: SO$_2$-N-ted (-5 kJ mol$^{-1}$) > SO$_2$-benzene (-22 kJ mol$^{-1}$) > SO$_2$-CH$_2$-ted (-61 kJ mol$^{-1}$) ≥ SO$_2$-O-bdc (-66 kJ mol$^{-1}$).

The initial configuration of SO$_2$ interacting with the nitrogen atom of the ted group evolves quite dramatically during the structural optimization, resulting in the complete desorption of the SO$_2$ molecule from the ted unit back into the MOF channel. The small adsorption energy observed (-5 kJ mol$^{-1}$) originates from the weak interactions of SO$_2$ with the surrounding linkers groups rather than with N. This situation has been expected since the crammed chemical environment of the ted group, and specifically the poor accessibility to the N atoms from the MOF-channel (see Figure 8d).

The model of the SO$_2$ molecule on the benzene site was investigated by both calculations (see Sec S11 in the Supporting information) and the experimental work of Taleb-Bendiab *et al*,[60] which show that the binding energy for this configuration is E$_b$(theory) = -22 KJ/mole and E$_b$(exp) = -18 KJ mol$^{-1}$, respectively, i.e. much less than for SO$_2$ in the other configurations SO$_2$-CH$_2$-ted and SO$_2$-O-bdc. Note that the calculations were done for Zn(bdc)(ted)$_{0.5}$, but very similar results are expected for the iso-structural Ni(bdc)(ted)$_{0.5}$ since the bonding does not involve the metal ions.

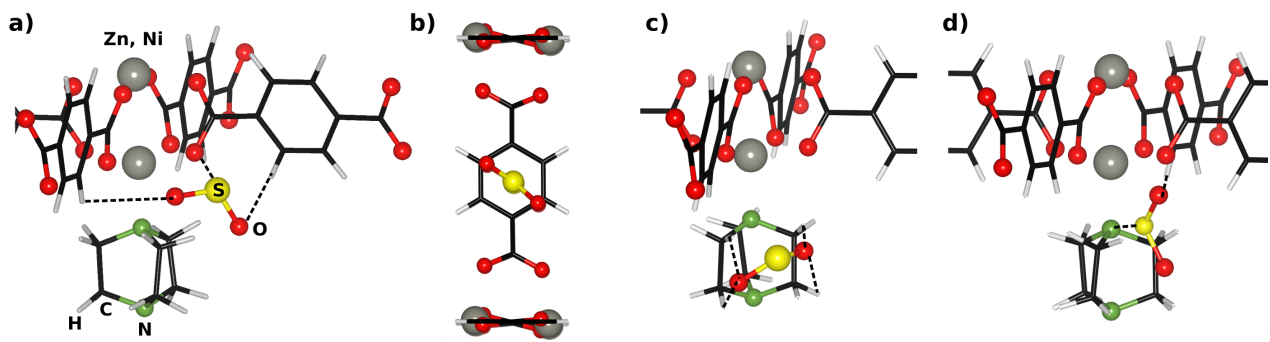

**Figure 8.** Snapshot of four initial adsorption configuration models used to investigate possible interactions of $SO_2$ in Zn, Ni(bdc)(ted)$_{0.5}$. a) $SO_2$-O-bdc, b) $SO_2$-benzene, c) $SO_2$-$CH_2$-ted, d) and $SO_2$-N-ted. Dashed lines represent the shorter contacts for $SO_2$ with the surrounding groups available on the linker unit

The Bader charge analysis of Table 2 suggests that in the cases of $SO_2$-O-bdc and $SO_2$-benzene, the sulfur atom of $SO_2$ always acts as acceptor, while the MOF-unit groups (i.e. benzene, or O) act as donors. This situation changes for the $SO_2$-$CH_2$-ted model where the hydrogen bonds are between O atoms of the $SO_2$ molecule and the H atoms of the $CH_2$ group of the ted unit (see Figure 8c).

**Table 2.** Bader charges (in units of the electronic charge) of sulfur in $SO_2$ and the adsorbing group (Grp.) before (as B) and after (as A) the adsorption. $\Delta e$ is the absolute difference between charges.

| Model | $SO_2$ | | | MOF-atom | | |
|---|---|---|---|---|---|---|
| | B | A | $\Delta e$ | Grp. | B | A | $\Delta e$ |
| -O-bdc | +2.85 | +1.73 | -1.11 | O | -1.19 | -1.08 | 0.10 |
| -$CH_2$-ted | +2.85 | +2.07 | -0.79 | H* | -0.06 | -0.09 | -0.03 |
| -benz. | +2.85 | +2.72 | -0.14 | C** | 0.05 | 0.05 | 0.00 |

*$SO_2$ is pointing with its oxygen atoms on H atoms of the $CH_2$-ted group see Fig. 8c.
**The charges on carbon atoms of the benzene were averaged.

The charge analysis of Table 2 shows that the "charge transfer" from the MOF towards the $SO_2$ molecule is correlated with the adsorption energy by assisting the electron-deficiency of sulfur. For instance, the sulfur atom in the $SO_2$-O-bdc and –$CH_2$-ted configurations almost changes its oxidation state during adsorption on the oxygen atom of the bdc group, suggesting a bonding mechanism stronger than typical physisorption. However, the binding energy (66 KJ/mole) is still in the range that is typically associated with physical adsorption. In contrast, $SO_2$ adsorption on the benzene rings leads to a small variation in the molecule and MOF charges, a clear indication of a weak van der Waals interaction that is consistent with lower calculated binding energy (~ 22 kJ mol$^{-1}$) and the shallow binding energy profile (see Sec S11 in the Supporting information). Note that the Bader analysis is an intuitive (but not unique) way of repartitioning the electron charge density, similar to other charge analyses, and therefore can only be used to draw qualitative conclusions.[61]

Initially, we postulated that the strongly bound species, characterized by the modes at 1242 and 1105 cm$^{-1}$ in Figure 3, 4 and 5, could be associated with the geometry shown in Figure. 8(c) with $SO_2$ located next to the ted $CH_2$ group. However the binding energy calculated from the $SO_2$-$CH_2$-ted model (-61 kJ mol$^{-1}$) is too low, even slightly less than that of the $SO_2$-O-bdc model in Figure. 8a, which is not consistent with the measured elevated desorption temperature (150 °C). The frequencies of $\nu_{as}$ and $\nu_s$ mode of $SO_2$ molecules in this configuration also exhibit a modest red-shift of -31 and -20 cm$^{-1}$, much less than the observed frequency shifts of the strongly bound species (see Sec 4.2). Therefore the theoretical calculations cannot shed light on the binding configuration of the strongly bound $SO_2$. The spectroscopic results in Figure 7 clearly show that this adsorption state is associated with the ted linker because its $CH_2$ stretching modes are significantly perturbed. Interestingly, the inclusion of this strongly bound species brings in significant modifications to carboxylate and $CH_x$ vibrational modes of the MOF (as shown in Sec 3.2, 3 and Figure S8). Further extensive additional theoretical and experimental studies necessary to fully understand this interaction are beyond the scope of this work. They will involve the preparation of more isotypical MOFs containing ted linkers, which is important to explore more binding geometries with different expected $SO_2$ interactions with the ted group.

In order to fully investigate the loading capabilities of M(bdc)(ted)$_{0.5}$ and the MOF structure perturbation, we have progressively adsorbed an increasing number of $SO_2$ molecules (1-8 molecules) until the available oxygen atoms are completely saturated. Here we explore the effect of loading only for the most stable adsorption case i.e. $SO_2$-O-bdc and $SO_2$-$CH_2$-ted. The MOF Zn(bdc)(ted)$_{0.5}$ structure displays 8 oxygen atoms originating from bdc linkers (see red atoms at the center of Figure 8b). In Table 3 we report the adsorption energies (with relative contribution) along with relevant bondlengths and structural parameters of the most relevant adsorption geometries, i.e. $SO_2$-O-bdc and $SO_2$-$CH_2$-ted.

Table 3 shows that $SO_2$ binds strongly with the oxygen atoms in the Ni and Zn(bdc)(ted)$_{0.5}$ structures, which is expected to affect the frequencies of $\nu$(COO) modes as observed in Figure 5. Figure 9 highlights the limited configurations surrounding the metal sites in M(bdc)(ted)$_{0.5}$ available for $SO_2$ adsorption. Figure 9 also shows the formation of hydrogen bonds between the oxygen atoms of $SO_2$ and the hydrogen on the benzene ring of the linkers. The magnitude of the distortion imposed by SO$_2$ on the MOF structure can be monitored by following the variation of the angle ϕ of Table 3. The large distortion imposes dramatic changes in the linker structure, affecting their vibrational frequencies as observed experimentally in the ring deformation mode σ$_{12}$ and other ring stretching mode ν$_{18a}$, ν$_{19b}$, ν$_{19a}$ shift in Figure 5.

**Table 3.** Adsorption energies ΔE (in kJ·mol$^{-1}$) of SO$_2$ in Zn, Ni (bdc)(ted)$_{0.5}$. ΔE is also corrected by the ZPE and thermal contribution ΔH at 298 K. Entropies, ΔS, are also reported. For the SO$_2$-O-bdc model S-O is the averaged intermolecular bond-length OOS—O-(bdc) in Å, whereas Δϕ is the variation of the torsion angle, in degrees, shown in Fig. 8.

| Loading | ΔE | ΔE$_{ZPE}$ | ΔH$_{298}$ | ΔS | S—O* | Δϕ |
|---|---|---|---|---|---|---|
| Model SO$_2$-CH$_2$-ted Zn(bdc)(ted)$_{0.5}$ | | | | | | |
| 1/8 | −61 | | | | 2.56$^a$ | − |
| Model SO$_2$-O-bdc Zn(bdc)(ted)$_{0.5}$ | | | | | | |
| 1/8 | −66 | −64 | −70 | −220 | 2.84 | −43 |
| 2/8 | −66 | −62 | −68 | −223 | 2.87 | −40 |
| 4/8 | −66 | −64 | −70 | −221 | 2.99 | −44 |
| 6/8 | −71 | −69 | −75 | −226 | 3.07 | −48 |
| 8/8 | −73 | −71 | −77 | −227 | 3.16 | −49 |
| Model SO$_2$-O-bdc Ni(bdc)(ted)$_{0.5}$ | | | | | | |
| 2/8 | −77 | −75 | −81 | −228 | 2.90 | -46 |

$^a$*S-O bond length becomes O-H for SO$_2$-CH$_2$-ted model (see Figure 8c).

In Figure 8(c), SO$_2$ also forms hydrogen bonds with the CH$_2$ group of the ted linker through its two oxygen atoms, as evidenced by the close distance (2.56 Å) between O and H atoms shown in Figure 8c and Table 3. This is an interesting binding geometry since the ted linkers provide a chelating site for the guest SO$_2$ molecules. The experimentally observed perturbation of the ρ(CH$_2$) mode shown in Figure 5 are consistent with the interaction between SO$_2$ and CH$_2$ (see Sec. 3.2).

The calculated adsorption energies are large for M(bdc)(ted)$_{0.5}$ and they increase with loading. However the calculated decrease of ΔE with the molecular loading could be an artifact of the increasing inter-molecular interactions, caused by the finite size of the cell adopted. For such reasons, our analysis of the absorption of SO$_2$ in Ni(bdc)(ted)$_{0.5}$ is only restricted to a representative model for two SO$_2$ per unit cell (2/8). The binding energies for this latter MOF are similar to those of Zn(bdc)(ted)$_{0.5}$.

The above adsorption energy calculations, combined with the later IR frequencies simulation, suggest that the infrared bands at 1326 and 1144 cm$^{-1}$ species could be due to SO$_2$ trapped in either SO$_2$-CH$_2$-ted or SO$_2$-O-bdc configurations since their adsorption energies and IR frequency shift as calculated in the next section are similar. The observation that the benzene ring deformation modes σ$_{12}$ and stretching modes of ν$_{18a}$, ν$_{19b}$, ν$_{19a}$ are affected by this weakly bound SO$_2$ additionally confirms the existence of the second configuration (SO$_2$-O-bdc) shown in Figure 8a.

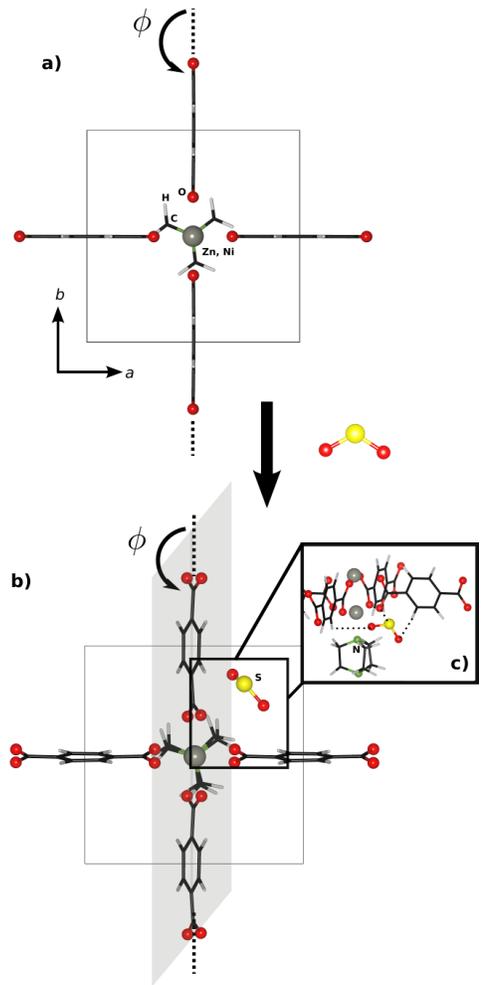

**Figure 9.** Top-view of M(bdc)(ted)$_{0.5}$ a) before and b) after the absorption of SO$_2$ molecules. c) Enlargement of the adsorption area for SO$_2$, where dashed lines show the interaction of S with O atoms (see bond length S—O in Table 2) of the MOF and hydrogen bonds.. The structural deformation upon SO$_2$ absorption is highlighted by the gray plane in a) and b) and the ϕ is the torsion angle

### 4.2 Simulation of IR spectra

We further discuss the effect of adsorption on the IR frequency modes of SO$_2$. The theoretical analysis concerns the vibrational modes of SO$_2$ and the MOF in gas-phase and once adsorbed into M(bdc)(ted)$_{0.5}$ (M = Zn, Ni), respectively, as reported in Table 4. The frequency calculations presented here are only performed for specific configurations: 2 SO$_2$/8 sites of loading for M(bdc)(ted)$_{0.5}$ (M = Zn, Ni) (see Table 2). The analysis focuses initially on the SO$_2$ frequency shifts observed when the first molecule is introduced in the MOF channel. According to the irreducible representation of SO$_2$, $C_{2v}$, there are three expected vibrational modes: i.e. two stretching modes (asymmetric and symmetric) and one bending mode.

From Table 4, it is clear that the computed IR frequencies of the SO$_2$ gas-phase are underestimated. To determine if such a discrepancy is due to the pseudo-potential (PS) employed, we performed a similar calculation using a "harder" PS, which increases the computational cost, but we observed only a marginal change of the absolute frequency values (about +20 cm$^{-1}$).

**Table 4.** Simulated IR vibrational frequencies (in cm$^{-1}$) for $SO_2$ in gas-phase and within M(bdc)(ted)$_{0.5}$ (M = Ni, Zn). Experimental frequencies value for the $SO_2$ gas-phase are also reported. In bracket we report the frequency shifts with respect to those calculated for the gas-phase.

|  | $SO_2$-gas | | M(bdc)(ted)$_{0.5}$ | | |
| --- | --- | --- | --- | --- | --- |
|  | Exp. | Cal. | Ni | Zn ($SO_2$-O-bdc) | Zn ($SO_2$-$CH_2$-ted) |
| $\nu_{as}$ | 1362$^a$ | 1256 | 1217 (-39) | 1224 (-32) | 1225 (-31) |
| $\nu_s$ | 1151$^a$ | 1073 | 1053 (-20) | 1057 (-16) | 1053 (-20) |
| $\delta$ | 497$^b$ | 498 | 491 (-7) | 501 (+3) | 497 (-1) |

$^a$This work (see Sec. 3.2).
$^b$Ref. 62

Nevertheless, the quantity of interest here is the frequency shift upon $SO_2$ adsorption within the MOF structure, which is more accurately determined than the absolute frequencies. Table 4 shows that the $SO_2$ vibrational modes experience a shift once in the M(bdc)(ted)$_{0.5}$ structure. In both $SO_2$-O-bdc and $SO_2$-$CH_2$-ted configurations, the molecular adsorption leads to similar red-shifts of the stretching modes, $\nu_3$ and $\nu_1$, as shown in Table 4, a clear indication that those bonds are weakened due to the $SO_2$ interaction with the MOF. The bending modes are not affected much after the adsorption of $SO_2$. Experimentally for Ni(bdc)(ted)$_{0.5}$ two sharp bands are observed at 1326 cm$^{-1}$ and 1144 cm$^{-1}$, respectively, corresponding to ~-36 cm$^{-1}$ and ~-7 cm$^{-1}$ red-shifts (from the unperturbed value at 1362 cm$^{-1}$ and 1151 cm$^{-1}$) of the asymmetric and symmetric S=O stretch modes of the $SO_2$ molecules (see Sec. 3.2). Although the absolute theoretical values underestimate the values of the frequencies observed experimentally, the red shifts are in good agreement with the experimental measurements (see Table 4). As a guide for future interpretation we report in Figure S14 the computed frequencies in form of the spectra, where only stretching modes are shown.

We also investigated the effect of the $SO_2$ adsorption on the vibrational modes of the MOF framework. Note that a full symmetry analysis of each mode associated with the MOF as well as a full description of each mode become extremely challenging; we therefore limit the discussion of Table 5 to the frequency modes that are detected experimentally (see Sec 3.2). Furthermore, the conclusions derived from the frequency calculations of the model $SO_2$-$CH_2$-ted segment of the Zn(bdc)(ted)$_{0.5}$ framework are assumed to carry over for Ni(bdc)(ted)$_{0.5}$ as well.

Table 5 shows that our calculations slightly overestimate the experimental CH stretching modes, while underestimate the COO asymmetric and symmetric frequencies (see Sec. 3.2). In general, the shift of the calculated vibrational modes is quite small. Most of vibrational modes investigated are strongly affected both in intensity and peak position by the adsorption of a strongly bound species with modes at 1242 and 1105 cm$^{-1}$ (see Sec 3.2), making it difficult to analyze the measured peak

**Table 5.** Simulated IR vibrational frequencies (in cm$^{-1}$) for $SO_2$ in gas-phase and within Zn(bdc)(ted)$_{0.5}$. In bracket we report the frequency shifts with respect to those calculated for the bare MOF. Grp. refers to the vibrating group, whereas $m$ for modes.

| Zn(bcd)ted$_{0.5}$ | | | |
| --- | --- | --- | --- |
| Grp. | $m$ | M(bcd)ted$_{0.5}$ | $SO_2$-O-bdc | $SO_2$-$CH_2$-ted |
| CH (bdc) | $\nu$ | 3053 | 3069 (+16) | 3066 (+13) |
| $CH_2$(ted) | $\nu_{as}$ | 3036 | 3040 (+4) | 3043 (+7) |
| $CH_2$ (ted) | $\nu_s$ | 2996 | 3001 (+5) | 3002 (+6) |
| COO(bdc) | $\nu_{as}$ | 1549 | 1561 (+12) | 1560 (+11) |
| COO(bdc) | $\delta$ | 849 | 846 (-3) | 848 (-1) |
| $CH_2$(ted) | $\rho$* | 821 | 829 (+8) | 823 (+2) |
| COO(bdc) | $\delta_{oop}$ | 788 | 788 (0) | 789 (+1) |
| Ni(bcd)ted$_{0.5}$ | | | |
| Grp. | $m$ | M(bcd)ted$_{0.5}$ | $SO_2$-O-bdc | |
| CH(bdc) | $\nu$ | 3050 | 3072 (+22) | |
| $CH_2$(ted) | $\nu_{as}$ | 3030 | 3040 (+10) | |
| $CH_2$ (ted) | $\nu_s$ | 2988 | 2991 (+3) | |
| COO(bdc) | $\nu_{as}$ | 1519 | 1530 (+11) | |
| COO(bdc) | $\delta$ | 863 | 860 (-3) | |
| $CH_2$ (ted) | $\rho$* | 829 | 835 (+6) | |
| COO(bdc) | $\delta_{oop}$ | 777 | 778 (+1) | |

*$\rho$ is the bending mode out of plane.

position change induced by the more weakly adsorbed $SO_2$ (with modes at 1326 and 1144 cm$^{-1}$). The differential spectra in Figure 5 show that the perturbations of bdc and ted modes are partially released upon desorption of this more weakly bound species, providing indirect evidence for $SO_2$ interaction with the functional groups of organic linkers.

## 5. Conclusion

In this work, we have shown that the metal organic framework Ni(bdc)(ted)$_{0.5}$ adsorbs a large quantity of polluting $SO_2$ gas [9.97 mol kg$^{-1}$] at 1.13 bar at room temperature. The high uptake capacity can be attributed to multiple interactions of $SO_2$ molecules within the framework units. These include the interaction of sulfur atoms with paddlewheel metal oxygen carbon units of the secondary building unit and oxygen atoms of $SO_2$ with the C-H, $CH_2$ of organic linkers, as observed by infrared spectroscopy and supported by *ab initio* DFT calculations. The IR data also point to another configuration of $SO_2$ that is more strongly bound to the organic linker ted, and can be removed at higher temperature (~ 150$^o$C) without destroying the frameworks structure. These findings are important to develop frameworks based on paddlewheel building units for the efficient capture of $SO_2$.

## ASSOCIATED CONTENT

**Supporting Information**. Mg-MOF-74 crystal structure and adsorption and desorption isotherm of $SO_2$, PXRD pattern of M(bdc)(ted)$_{0.5}$ and TG profiles, Infrared spectra of Ni(bdc)(ted)$_{0.5}$ and its vibrational assignments, different gas adsorption capacities in Ni(bdc)(ted)$_{0.5}$, IR spectra of $SO_2$ adsorption into Zn(bdc)(ted)$_{0.5}$. Co-adsorption experiment of $CO_2$ and $SO_2$, binding energy curve, and simulated IR spectra. This material is available free of charge via the Internet at http://pubs.acs.org."

# AUTHOR INFORMATION

## Corresponding Author

* Phone 1-972-883-5751; e-mail: chabal@utdallas.edu.

## ACKNOWLEDGMENT

The synthesis, spectroscopic characterization and modeling work performed at Rutgers, UT Dallas, and Wake Forest was supported in its totality by the Department of Energy, Basic Energy Sciences, division of Materials Sciences and Engineering (DOE grant No. DE-FG02-08ER46491). The isotherm measurements, performed at PNNL, were supported by the U.S. Department of Energy, Office of Basic Energy Sciences, Division of Materials Sciences and Engineering (Award KC020105-FWP12152). Pacific Northwest National Laboratory is operated by Battelle for the U.S. Department of Energy under Contract DE-AC05-76RL01830.